\documentclass{mem}
\usepackage{natbib}
\usepackage{txfonts}
\usepackage{balance}
\usepackage{graphicx}
\usepackage[usenames,dvipsnames]{color}
\usepackage{colortbl,xcolor}
\usepackage{multirow}
\usepackage[a4paper,breaklinks,dvipdfm]{hyperref}
\idline{75}{282}

\newcommand{\swift}{\textit{Swift}}
\newcommand{\theseus}{\textit{THESEUS}}

\begin{document}
\def\teff{$T\rm_{eff }$}
\def\kms{$\mathrm {km s}^{-1}$}

\title{
GRB follow-up and science with THESEUS/IRT
}

   \subtitle{}

\author{
A. \,Rossi\inst{1} 
\and G. \,Stratta\inst{1,2,3} 
\and E. \,Maiorano\inst{1}
\and L. \,Amati\inst{1} 
\and L. \,Nicastro\inst{1} 
\and E. \,Palazzi\inst{1}  
}

\institute{
INAF-OAS Bologna, Area della Ricerca CNR, via Gobetti 101, I--40129 Bologna, Italy
\email{a.rossi@iasfbo.inaf.it}
\and
Urbino University, Via S. Chiara 27, 61029, Urbino (PU), Italy
\and
INFN-Firenze, via G. Sansone 1, 50019 Sesto Fiorentino (FI), Italy
}

\authorrunning{Rossi}

\titlerunning{GRB science with THESEUS}

\abstract{ 
The aim of the space mission concept \theseus~is to
continue to collect and study the GRB events like \swift.
It will allow us to study the early Universe. Moreover, it will 
offer us to study with unprecedented sensitivity GRB emission
and to measure the redshift for the bursts with $z>5$.
In this work, we investigate the advantages of a 
optical and near-infrared telescope mounted on the same satellite that is triggered
by the GRB like \theseus/IRT. Afterwards, we investigate the possible future developments 
in the GRB science, first for the prompt phase and the for afterglow phase.
We find that more than half of the sources detected by \theseus, and will never be visible
from a a ground-based telescope. Moreover, only $\sim50\%$ of all observable sources are visible within one hour, 
i.e. $<30\%$ of all \theseus~transient sources. A higher number of observable sources can only be achieved with a network
of telescopes.
\theseus~will permit to detect the NIR prompt phase of the longest GRBs,
increasing the number of events studied from gamma-rays to the near-infrared 
from a handful of events studied up to now to $\gtrsim10$ GRBs per year.

\keywords{Gamma rays: bursts}
}
\maketitle{}

\section{Introduction}

After 50 years since their discovery, Gamma-ray bursts (GRBs) 
are still one of the most fascinating research fields in Astrophysics. 
Indeed, they are the most energetic gamma-ray emitter known, and a ultra-relativistic 
laboratory for high energy physics, high-redshift environment, massive-star formation and, cosmology.

After the discovery that GRBs are not local but cosmological explosions 
\citep[thanks to the \emph{Bepposax} space observatory][]{Boella1997b} 
of massive stars \citep[e.g.,][]{Hjorth2003a}, in the last 12 years, 
\swift~\citep{Gehrels2004} allowed us to further 
improve our understanding of GRB phenomena, thanks to 
\swift's rapid and autonomous slewing capabilities, in combination with its
highly sensitive X-ray telescope (XRT; \citealt{Burrows2005a}) as well as its
optical/UV telescope (UVOT; \citealt{Roming2005a}).  Today, about 50 to 70
GRB optical afterglows can be localized annually by \swift, with 30 to 40 having redshifts
determined mostly by ground-based observatories.

However, \swift~is far beyond its planned life, and yet many questions are open:
How the GRB engine exactly works?
Why the GRB afterglow emission is so different from case to case?
Is there a unique class of GRB progenitor and how is the influence of the environment in their formation?
Are long GRBs progenitors good tracers of star formations? 
In order to answer these and many other questions further studies are in order.
 
The space mission concept \theseus~\citep{Amati2017a} aims not only 
to continue to collect and study the GRB events like \swift, 
but to measure directly the redshift of GRBs at $z>$5.
Thus, it will allow us to make use of GRBs to study the early Universe, in particular 
star formation rate and metallicity evolution of the interstellar 
and intra-galactic medium up to redshift $\sim10$, signatures of Pop III stars, sources and physics of re-ionization,
and the faint end of the galaxy luminosity function. 
\theseus~will also provide unprecedented capability 
to localize the electromagnetic counterparts of gravitational radiation.

The number of expected GRBs triggers per year from the \theseus~high-energy instruments 
(SXI and XGS) varies from 387 to 870. 
By taking the average value, one could expect about 2 triggers per day. Following-up all
the triggers systematically and providing the associated redshift estimate is the key in order to
be able to have enough high-redshift GRBs to fulfil the mission requirements. The near-infrared (NIR) telescope IRT on
\theseus~gives access to the early afterglow and the late prompt phase of GRBs in some
cases, a poorly studied interval so far.

In the following sections, we will first investigate the advantages of a 
optical and near-infared telescope mounted on the same satellite that is triggered
by the GRB. Afterwards, we investigate the possible future developments 
in the GRB science, first for the prompt phase and the for afterglow phase.

\begin{figure}[htbp!]
\resizebox{\hsize}{!}{\includegraphics[clip=true]{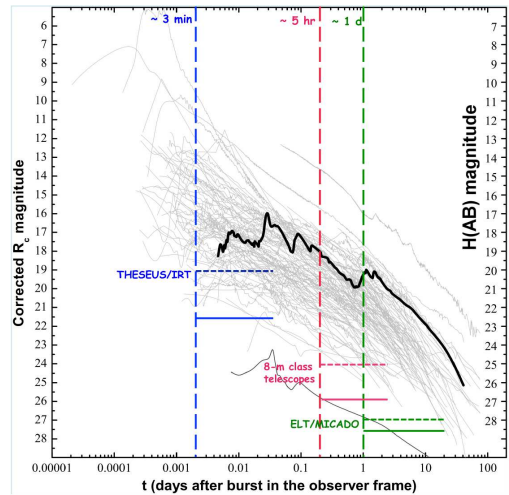}}
\caption{\footnotesize
Observed R band lightcurves of long GRBs adapted from \citealt{Kann2017a}. The left axis indicates the H$_{AB}$ magnitudes, obtained assuming standard afterglow color (see text). Highlighted is the ultra long GRB 111209A and the extremely extinguished GRB 130925A (bottom curve). Data is corrected for Galactic extinction. Adapted from Kann et al. (2017). We also show in the figure with blue lines the \theseus/IRT sensitivity, the VLT FORS and X-Shooter sensitivity in red and, the ELT sensitivity in green (dashed line spectroscopy, solid line imaging).
}
\label{fig:kannlc}
\end{figure}

\section{Comparison between \theseus/IRT and other facilities}

The sample of long GRB lightcurves collected in the studies of Kann et al., 
\citep[e.g.,][]{Kann2017a,Kann2010a}, showed that almost all ($>90\%$) the afterglows of GRBs observed to date within few minutes are brighter than $R\sim21$. 
Assuming an optical to NIR spectral slope of $\beta=0.7$, the $R-H$ color is $\sim$0.7 mag (AB system).
In this way, we can use Figure \ref{fig:kannlc} and conclude that
 that almost all ($>90\%$)
the NIR afterglows of GRBs observed to date within few minutes after the trigger are brighter
than $H_{AB}\sim$20, and they fade rapidly to $H_{AB}>24$ within few hours. 

Note that \swift/XRT, with an orbit and pointing constraints similar to what is planned 
for \theseus/IRT, is capable to follow-up 
$>80\%$ of GRBs. \swift/UVOT detects about $40\%$ of afterglows \citep{Roming2009a}, 
and it is limited by its size and the fact to operate in UV/optical, thus misses the
most extinguished or $z>4$ GRBs, all aspects that IRT will overcome. Even with these
limitations, \swift/UVOT behaves better than any robotic ground-based telescope that recover
only $20-30\%$ of the afterglows (\texttt{\small http://www.mpe.mpg.de/$\sim$jcg/grbgen.html} and \citealt{Greiner2011}). 
Even if one considers the worst scenario of a GRBs detected close
to the border of SXI FOV, IRT would be capable to detect all afterglows known to date and
measure a photometric redshift for $\sim90\%$ of the cases ($H_{AB}>19.5$), starting observing
in LRS mode 20 min after the trigger \citep{Amati2017a,Goetz2018a}. This can be compared to a dedicated follow-up program
on a 2m optical/NIR instrument like GROND mounted on the 2.2m ESO (La SIlla Observatory),
which is capable to follow-up and detect $<20\%$ GRBs within 30 minutes and $30\%$ within 4 hours \citep{Greiner2011},
in agreement with what we find in section \ref{sec:wt}.
Note that, assuming that $z>6$ afterglows have IR luminosity and evolution similar to those
observed up to date, $>50\%$ GRB will have $H_{AB}<19.5$ at 5 min after the trigger, thus they
can be detected by IRT using LRS spectra and a photo-z can be measured.
For the performances of IRT and the observing strategy of IRT see \citet{Amati2017a} and \citet{Goetz2018a}. 
Finally, space
observations are also not affected by sky absorption, which, especially at nIR wavelengths,
makes large spectral intervals completely not accessible from ground thus affecting the
spectral studies.

\begin{figure}[ht!]
\resizebox{\hsize}{!}{\includegraphics[clip=true]{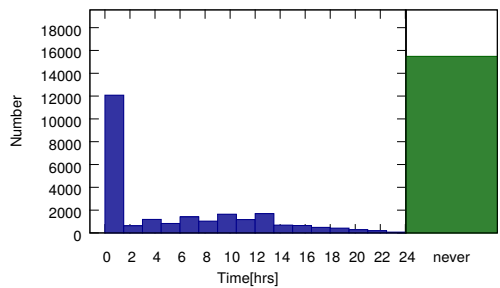}}
\caption{\footnotesize
{\it Left}: Distribution of the time on source TOS, i.e. the time needed before a transient source is visible from a ground observatory, in this case 
ESO Paranal Observatory (Chile).
{\it Right}: Sources that are never observable from Paranal. 
}
\label{fig:wt}
\end{figure}
\begin{figure}[ht!]
\resizebox{\hsize}{!}{\includegraphics[clip=true]{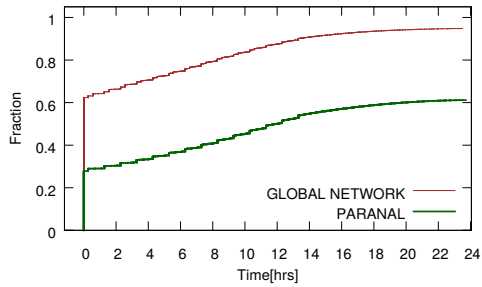}}
\caption{\footnotesize
Cumulative distribution of the time on source TOS. Observations from ESO/Paranal are in green. 
Observations from a global network of telescopes (red).
}
\label{fig:wt2}
\end{figure}

\section{Ground-based follow-up of \theseus~triggers \label{sec:wt}}

The ground-based
telescopes capable to reach a NIR sensitivity similar or better than IRT are larger than 2 m and are concentrated between
Hawaii and Canary Islands in the northern hemisphere and between Chile and South Africa in
the southern hemisphere.
 Thus, more than $\sim30/50\%$ of the northern/southern hemisphere is
not visible for several hours from ground-based telescopes. At this time, half of the afterglows
are bright enough only for the few 4~m or larger telescopes. 
Moreover, these telescopes are
also limited by the impossibility to override other observations (i.e. time critical program
running), which will likely happen (especially for ELT class telescopes and JWST) given that
we expect to follow up more than 1 GRB per day. 

In the following, we want to investigate the time needed by a 
ground-based observatory to be on source, i.e., the time on source (in the following: TOS).
It is important to first notice that
the IRT will be mounted  on the same satellite where is the instrument devoted 
to be triggered by the transient. Thus, all triggers will be observed by IRT.
The question, then, is how many triggers followed-up by IRT within a time interval 
(90 minutes, i.e., the \theseus~orbital period),  
can also be followed up by a ground-based observatory, and what is the TOS.

The ingredients to take in account depends from the observability of the transient which is given by: 
i) the trigger time, given in UT;
ii) the time interval when the transient is above a given elevation limit 
(we have chosen 30$^\circ$ as this is an usual limit in many observatories); it depends from the position on sky of the transient
and the coordinates of the observatory;
iii) the start, end and length of the astronomical night at the observatory site (sun elevation $>-18^\circ$), 
which depends from the coordinates of the observatory site.
The time spent before a transient is observable is the TOS.

Using the above ingredients, we have
generated several thousands of sources triggered by a space observatory like \theseus~and calculated the TOS for a facility based on ground.
The sources are uniformly distributed in the sky and their coordinates are in the intervals $0^\circ<RA<360^\circ$,$-60^\circ<DEC<60^\circ$. From these, we removed the sources too close in projection to the sun, i.e., those closer than $40^\circ$. 
This should be approximately the region on sky that will be more often covered by \theseus, i.e. where it is more likely that
a transient source will be. This is justified by the $\pm60^\circ$ region on sky covered by SXI.
 We caution that this is just an approximation of the real case, because \theseus~real orbit will not be perfectly equatorial 
 and its pointing will be constrained strongly by the Sun. 
We generated $\sim400$ sources positions, and repeated the process for a observing time step of 
90 minutes (approximately the duration of THESEUS orbit), 
during 24 hours and 4 times a year 
(i.e., summer and winter equinoxes and solstices), for a total of $\sim40000$ sources.
As typical observatory site, we have chosen the ESO/VLT at Paranal, which is good also for other international 
facilities in Chile (e.g., ELT, LSST, Magellan, GROND).
Figure~\ref{fig:wt} shows the result of this simulation. Slightly less than half of the sources 
will never be visible from the site, and only $\sim30\%$ of all sources are visible within one hour.

In order to maximize the number of triggers observable from ground, one would require
a high number of robotic telescopes with NIR cameras in order to cover the entire Earth
latitudes (at least every 45$^{\circ}$ in latitude) to cope with the day/night limitations and weather
uncertaintie. Those robotic telescopes would need to be of the 2-4 m class in order to have
an equivalent sensitivity to the IRT in space. Those facilities are not available to date, and it
would require a complex international strategy to put them in place.
We simulated a network of 6 sites which are already home of large telescopes and at very different longitudes to
cover most of the world (ESO/Paranal, Hawaii, California, Arizona, Canary Islands, Xinglong/China).
Figure \ref{fig:wt2} shows the cumulative distribution of TOS time for the global network and, for comparison, that for PARANAL.
The global network allows to observe almost all IRT sources within 1 day, but only $\sim60\%$ of IRT sources within one hour. However, it doubles what is achieved from a single observatory (Paranal).

\begin{figure}[t!]
\resizebox{\hsize}{!}{\includegraphics[clip=true]{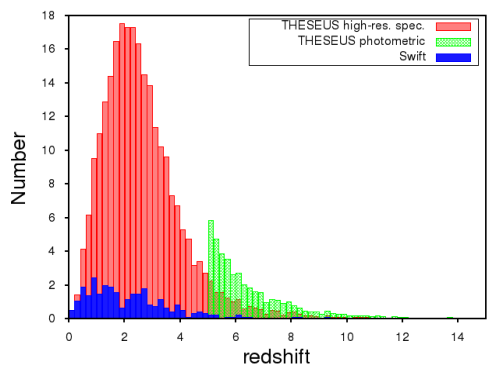}}
\caption{\footnotesize
The redshift distribution of \theseus~GRBs during a 5 yr
mission lifetime compared to the actual distribution of \swift~GRBs
(blue) during the same period. GRBs with a photometric redshift are in
green, and those with a spectroscopic redshift are in red.
}
\label{fig:grbs}
\end{figure}
\begin{figure}[t!]
\resizebox{\hsize}{!}{\includegraphics[clip=true]{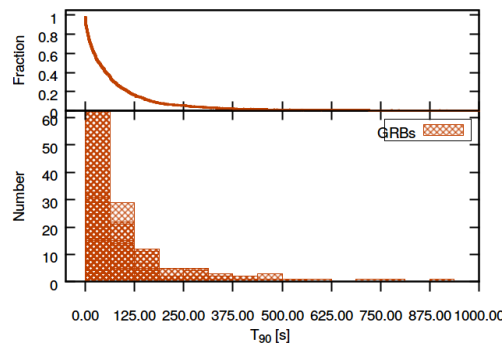}}
\caption{\footnotesize
Distribution of the T90 duration of \swift/XRT GRBs.
Note that less than 10$\%$ of all GRBs have a duration longer than 300 seconds.
}
\label{fig:t90}
\end{figure}

\section{Multiwavelength prompt emission}
\begin{figure}[t!]
\resizebox{\hsize}{!}{\includegraphics[clip=true]{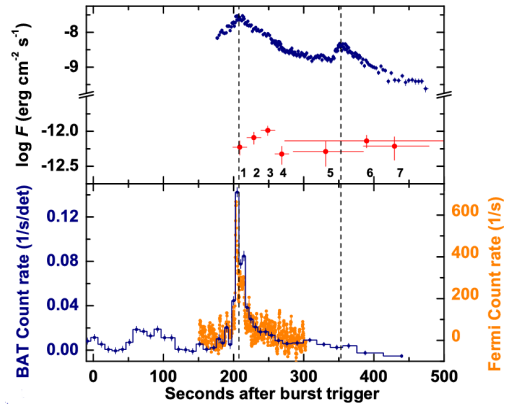}}
\caption{\footnotesize
Temporal evolution of the optical (red, top panel), X-ray (blue, top panel), and gamma-ray (bottom panel) 
emission of GRB 080928. The dashed vertical lines indicate the peak
times of the two X-ray flares 
\citep[adapted from][]{Rossi2011a}.
}
\label{fig:prompt}
\end{figure}

IRT will be capable to detect the NIR prompt phase of the longest GRBs ($>300$ sec),
increasing the number of prompt events studied from gamma-rays to NIR to 10 to 20 GRBs
per year (Fig. \ref{fig:grbs} and Fig. \ref{fig:t90}). Today, they are limited to an handful of cases (Fig. \ref{fig:prompt}). 

Right now, thanks to \swift/UVOT and ground-based optical/NIR telescopes dedicated to the rapid follow-up of GRB
afterglows, it is possible to simultaneously follow-up the
prompt emission from optical to X-ray to Gamma-rays. These
observations allow us not only to better constrain the spectral energy distribution from optical to Gamma-rays and their
lightcurves, testing the standard model, but also to test the
nature of the central engine. Indeed, the observer may see
simultaneously photons that have been emitted in different
times and regions of the flow, and also with different physical origin, e.g., synchrotron or synchrotron self-compton
emission. However, there are only a few tens of bursts in
12 years of \swift~activity that could be long and bright enough
to be detected in optical \citep{Levan2014a}. Indeed, up to
now only 6 events have been studied in such detail \citep{Bloom2009a,Rossi2011a,Stratta2013a,Elliott2014a,Troja2017a}, 
and while they probe that standard fireball
model can explain the observations, in some cases the optical and high-energy emission seems unrelated, or require a
more complex modelling of the jet structure (see Fig.~\ref{fig:prompt}). Moreover, these
observations have been performed in the optical, which is
more affected by foreground and host line of sight dust extinction. 
With \theseus/IRT capability of starting to obtain
the first images within the first 5 min from the trigger, it will
be possible to detect optical prompt emission for the longest
GRBs, roughly 10 to 20 GRBs per year. This will dramatically increase the number of events to study, and allow us
to statistically explore several models, shedding light on the structure of the jet during its first
phases.

\section{Complete samples}
In order to study properties of GRBs, their X-ray and optical/NIR afterglows, and their hosts is important to handle
a sample that is not biased towards classes (i.e. a complete sample), because of limitation during the observations. 
Until now, several GRB complete samples have being created \citep[e.g.,][]{Greiner2011,Salvaterra2012a,Perley2016a}. 
In addition to these, other tools to overcome the
problems of unbiased distributions with robust and sophysticated statistical techniques have already been successfully
applied to GRB prompt and afterglow emission \citep[e.g.,][]{Dainotti2013a,Dainotti2015a}. 
The capability of \theseus~to detect
most afterglows in the IR, excluding the few highly extinguished or at extreme redshift ($<10\%$), will allow us for the
first time to build a complete sample of GRB afterglows observed in X-rays and IR. The fact that \theseus~will not be
limited by weather conditions and visibility constraints, but
only from pointing limitation and foreground Galactic extinction, is a strong advantage in respect to ground-based 
facilities dedicated to GRB follow-up (e.g. RATIR, GROND,REM).

\section{Chances of misidentification}
Minimizing the TOS will minimize the chances of wrong association 
of host galaxies. This is of fundamental importance in the case of short GRBs (sGRBs),
which afterglow is much fainter then the long one \citep[e.g.,][]{Kann2011a} and often only a X-ray position is available.
This is further complicated by the fact that the sGRB progenitor can travel several Kpc away from its birth site before to explode
\citep[e.g.,][]{Berger2014a}. 
In the case of long GRBs (lGRBs), this often happens for those that are called dark GRBs, i.e. bursts which 
optical/NIR afterglow is dimmed by a combination of dust extinction and high redshift \citep[e.g.,][]{Rossi2012a}.
The recent finding of \cite{Perley2017a} that the host of dark GRB 020819B was not a foreground spiral 
galaxy but a background high redshift galaxy is a clear example. The wrong host was 
often erroneously used as an unique example of GRBs host with high metallicity at low redshift, 
leading to a puzzling contradicting conclusions about the condition necessary to the progenitor of lGRBs to form.
Another puzzling case is lGRB 050219A, which apparently exploded within a early-type galaxy, a unique case
among lGRBs \citep{Rossi2014a}, but the real host may be a smaller companion or background galaxy.
More in general, \cite{Hjorth2012a} estimates that up to $12\%$ of lGRB-host association may be wrong.
To minimize tha chances of a wrong host association, it is important to search for the host 
within a region that has to be as small as possible. Optical/NIR localization are better than X-rays, 
especially in the case of \theseus~(less than one arcsecond compare to several arcseconds).
It is therefore important to minimize the TOS, 
and thus to detect the transient when is brighter, especially in the NIR, which is less affected by dust extinction 
and not affected by Lyman absorption up to redshift $z\gtrsim12$.

\section{Optical/IR afterglow detection with IRT}

As shown in Figure\ref{fig:kannlc}, within the first hour all
known optical afterglows have $R<22$. A classical optical 
afterglow has a spectral slope $\beta\sim1$, which translates in a color
$R-H \sim1$ mag (AB system), thus within the first
hour all known afterglows have $H_{AB}<21$. 
IRT will observe optical afterglows longer than 30 min within 1 hour from the
trigger \citep{Goetz2018a}, reaching $H_{AB}\sim20.6$. The optical/NIR imager GROND,
reaching 1 mag fainter limits only, has been able to detect
$\sim90\%$ of all GRBs detected by \swift~within 4 hours from
the trigger \citep{Greiner2011}. Note that the host extinction
will mostly have a negligible effect, with only a few cases
($\sim 10\%$) with $AV >0.5$, which will noticibly dim ($>1$ mag) the
observed NIR afterglow when the redshift is $z>4$, and still
obtaining a detection rate of $\sim90\%$. However, at these redshift 
dusty environments are less common, because dust did
not have the time to accumulate in the star forming regions.
Notably, the higher rate of \theseus~GRBs will allow us to
better understand the shape of dust extinction curve at high
redshift which is now unexplored, the presence of 2175~\AA\ 
absorption in GRB SEDs in high redshift environments and
to test different models for dust grains.

Compared to today, the larger number of \theseus~GRBs and the more
sensitive spectra observed with XGIS will allow us to better 
understand the nature of the afterglow and of the central
engine of GRBs. The study of the optical/NIR and X-ray
afterglows unveils the properties of the environment. It is
well known that the circum-burst density profile influences
the shape of the GRB light curves and spectra \citep{Racusin2009a}
 distinguishing by ISM and wind environments \citep[e.g.,][]{Schulze2011}. 
 Moreover, dust
and gas on the line of sight dim the optical/NIR and X-ray
afterglows. Their
systematic study will unveil the properties of the environment where GRBs explode  \citep[e.g.,][]{Schady2012a}. 

The X-ray lightcurves of GRBs are often show a slow fading plateau phase that can persist even longer than
$10^4$ sec \citep[e.g.,][]{DePasquale2016a}. This plateau phase is difficult to explain within the collapsar scenario, because
it requires a long activity of the central engine.
Alternatively, the necessary energy may come from the spin-down activity
of a magnetar formed during the collapse 
\citep[e.g.,][]{ZhangMeszaros2001a}. 
To complicate this view, X-ray flares
(not visible in gamma-rays), probably due to late emission,
indicate that the central engine is still active.
Unfortunately, up to today the multi-wavelength study of these features
has been limited by the different time coverage of X-ray
and optical/NIR observations and sensitivity to the late afterglows. 
\theseus~will likely solve this problem thanks
to the simultaneous observations of optical/NIR and X-ray
afterglows.


\begin{acknowledgements}
I acknowledge support from the INAF project "LBT premiale 2013".
\end{acknowledgements}

\bibliographystyle{aa}
\bibliography{Rossi_grbscienceTheseus}

\begin{thebibliography}{32}
\expandafter\ifx\csname natexlab\endcsname\relax\def\natexlab#1{#1}\fi

\bibitem[{{Amati} {et~al.}(2017){Amati}, {O'Brien}, {Goetz}, {Bozzo}, {Tenzer},
  {Frontera}, {Ghirlanda}, {Labanti}, {Osborne}, {Stratta}, {Tanvir},
  {Willingale}, {Attina}, {Campana}, {Castro-Tirado}, {Contini}, {Fuschino},
  {Gomboc}, {Hudec}, {Orleanski}, {Renotte}, {Rodic}, {Bagoly}, {Blain},
  {Callanan}, {Covino}, {Ferrara}, {Le Floch}, {Marisaldi}, {Mereghetti},
  {Rosati}, {Vacchi}, {D'Avanzo}, {Giommi}, {Gomboc}, {Piranomonte}, {Piro},
  {Reglero}, {Rossi}, {Santangelo}, {Salvaterra}, {Tagliaferri}, {Vergani},
  {Vinciguerra}, {Briggs}, {Campolongo}, {Ciolfi}, {Connaughton}, {Cordier},
  {Morelli}, {Orlandini}, {Adami}, {Argan}, {Atteia}, {Auricchio}, {Balazs},
  {Baldazzi}, {Basa}, {Basak}, {Bellutti}, {Bernardini}, {Bertuccio}, {Braga},
  {Branchesi}, {Brandt}, {Brocato}, {Budtz-Jorgensen}, {Bulgarelli}, {Burderi},
  {Camp}, {Capozziello}, {Caruana}, {Casella}, {Cenko}, {Chardonnet}, {Ciardi},
  {Colafrancesco}, {Dainotti}, {D'Elia}, {De Martino}, {De Pasquale}, {Del
  Monte}, {Della Valle}, {Drago}, {Evangelista}, {Feroci}, {Finelli},
  {Fiorini}, {Fynbo}, {Gal-Yam}, {Gendre}, {Ghisellini}, {Grado}, {Guidorzi},
  {Hafizi}, {Hanlon}, {Hjorth}, {Izzo}, {Kiss}, {Kumar}, {Kuvvetli}, {Lavagna},
  {Li}, {Longo}, {Lyutikov}, {Maio}, {Maiorano}, {Malcovati}, {Malesani},
  {Margutti}, {Martin-Carrillo}, {Masetti}, {McBreen}, {Mignani}, {Morgante},
  {Mundell}, {Nargaard-Nielsen}, {Nicastro}, {Palazzi}, {Paltani}, {Panessa},
  {Pareschi}, {Pe'er}, {Penacchioni}, {Pian}, {Piedipalumbo}, {Piran}, {Rauw},
  {Razzano}, {Read}, {Rezzolla}, {Romano}, {Ruffini}, {Savaglio}, {Sguera},
  {Schady}, {Skidmore}, {Song}, {Stanway}, {Starling}, {Topinka}, {Troja}, {van
  Putten}, {Vanzella}, {Vercellone}, {Wilson-Hodge}, {Yonetoku}, {Zampa},
  {Zampa}, {Zhang}, {Zhang}, {Zhang}, {Zhang}, {Antonelli}, {Bianco}, {Boci},
  {Boer}, {Botticella}, {Boulade}, {Butler}, {Campana}, {Capitanio}, {Celotti},
  {Chen}, {Colpi}, {Comastri}, {Cuby}, {Dadina}, {De Luca}, {Dong}, {Ettori},
  {Gandhi}, {Geza}, {Greiner}, {Guiriec}, {Harms}, {Hernanz}, {Hornstrup},
  {Hutchinson}, {Israel}, {Jonker}, {Kaneko}, {Kawai}, {Wiersema}, {Korpela},
  {Lebrun}, {Lu}, {MacFadyen}, {Malaguti}, {Maraschi}, {Melandri}, {Modjaz},
  {Morris}, {Omodei}, {Paizis}, {Pata}, {Petrosian}, {Rachevski}, {Rhoads},
  {Ryde}, {Sabau-Graziati}, {Shigehiro}, {Sims}, {Soomin}, {Szecsi}, {Urata},
  {Uslenghi}, {Valenziano}, {Vianello}, {Vojtech}, {Watson}, \&
  {Zicha}}]{Amati2017a}
{Amati}, L., {O'Brien}, P., {Goetz}, D., {et~al.} 2017, ArXiv e-prints

\bibitem[{{Berger}(2014)}]{Berger2014a}
{Berger}, E. 2014, \araa, 52, 43

\bibitem[{{Bloom} {et~al.}(2009){Bloom}, {Perley}, {Li}, {Butler}, {Miller},
  {Kocevski}, {Kann}, {Foley}, {Chen}, {Filippenko}, {Starr}, {Macomber},
  {Prochaska}, {Chornock}, {Poznanski}, {Klose}, {Skrutskie}, {Lopez}, {Hall},
  {Glazebrook}, \& {Blake}}]{Bloom2009a}
{Bloom}, J.~S., {Perley}, D.~A., {Li}, W., {et~al.} 2009, \apj, 691, 723

\bibitem[{{Boella} {et~al.}(1997){Boella}, {Chiappetti}, {Conti}, {Cusumano},
  {del Sordo}, {La Rosa}, {Maccarone}, {Mineo}, {Molendi}, {Re}, {Sacco}, \&
  {Tripiciano}}]{Boella1997b}
{Boella}, G., {Chiappetti}, L., {Conti}, G., {et~al.} 1997, \aaps, 122, 327

\bibitem[{{Burrows} {et~al.}(2005){Burrows}, {Hill}, {Nousek}, {Kennea},
  {Wells}, {Osborne}, {Abbey}, {Beardmore}, {Mukerjee}, {Short}, {Chincarini},
  {Campana}, {Citterio}, {Moretti}, {Pagani}, {Tagliaferri}, {Giommi},
  {Capalbi}, {Tamburelli}, {Angelini}, {Cusumano}, {Br{\"a}uninger}, {Burkert},
  \& {Hartner}}]{Burrows2005a}
{Burrows}, D.~N., {Hill}, J.~E., {Nousek}, J.~A., {et~al.} 2005, Space Sci.
  Rev., 120, 165

\bibitem[{{Dainotti} {et~al.}(2015){Dainotti}, {Petrosian}, {Willingale},
  {O'Brien}, {Ostrowski}, \& {Nagataki}}]{Dainotti2015a}
{Dainotti}, M., {Petrosian}, V., {Willingale}, R., {et~al.} 2015, \mnras, 451,
  3898

\bibitem[{{Dainotti} {et~al.}(2013){Dainotti}, {Petrosian}, {Singal}, \&
  {Ostrowski}}]{Dainotti2013a}
{Dainotti}, M.~G., {Petrosian}, V., {Singal}, J., \& {Ostrowski}, M. 2013,
  \apj, 774, 157

\bibitem[{{De Pasquale} {et~al.}(2016){De Pasquale}, {Oates}, {Racusin},
  {Kann}, {Zhang}, {Pozanenko}, {Volnova}, {Trotter}, {Frank}, {Cucchiara},
  {Troja}, {Sbarufatti}, {Butler}, {Schulze}, {Cano}, {Page}, {Castro-Tirado},
  {Gorosabel}, {Lien}, {Fox}, {Littlejohns}, {Bloom}, {Prochaska}, {de Diego},
  {Gonzalez}, {Richer}, {Rom{\'a}n-Z{\'u}{\~n}iga}, {Watson}, {Gehrels},
  {Moseley}, {Kutyrev}, {Zane}, {Hoette}, {Russell}, {Rumyantsev}, {Klunko},
  {Burkhonov}, {Breeveld}, {Reichart}, \& {Haislip}}]{DePasquale2016a}
{De Pasquale}, M., {Oates}, S.~R., {Racusin}, J.~L., {et~al.} 2016, \mnras,
  455, 1027

\bibitem[{{Elliott} {et~al.}(2014){Elliott}, {Yu}, {Schmidl}, {Greiner},
  {Gruber}, {Oates}, {Kobayashi}, {Zhang}, {Cummings}, {Filgas}, {Gehrels},
  {Grupe}, {Kann}, {Klose}, {Kr{\"u}hler}, {Nicuesa Guelbenzu}, {Rau}, {Rossi},
  {Siegel}, {Schady}, {Sudilovsky}, {Tanga}, \& {Varela}}]{Elliott2014a}
{Elliott}, J., {Yu}, H.-F., {Schmidl}, S., {et~al.} 2014, \aap, 562, A100

\bibitem[{{Gehrels} {et~al.}(2004){Gehrels}, {Chincarini}, {Giommi}, {Mason},
  {Nousek}, {Wells}, {White}, {Barthelmy}, {Burrows}, {Cominsky}, {Hurley},
  {Marshall}, {M{\'e}sz{\'a}ros}, {Roming}, {Angelini}, {Barbier}, {Belloni},
  {Campana}, {Caraveo}, {Chester}, {Citterio}, {Cline}, {Cropper}, {Cummings},
  {Dean}, {Feigelson}, {Fenimore}, {Frail}, {Fruchter}, {Garmire}, {Gendreau},
  {Ghisellini}, {Greiner}, {Hill}, {Hunsberger}, {Krimm}, {Kulkarni}, {Kumar},
  {Lebrun}, {Lloyd-Ronning}, {Markwardt}, {Mattson}, {Mushotzky}, {Norris},
  {Osborne}, {Paczynski}, {Palmer}, {Park}, {Parsons}, {Paul}, {Rees},
  {Reynolds}, {Rhoads}, {Sasseen}, {Schaefer}, {Short}, {Smale}, {Smith},
  {Stella}, {Tagliaferri}, {Takahashi}, {Tashiro}, {Townsley}, {Tueller},
  {Turner}, {Vietri}, {Voges}, {Ward}, {Willingale}, {Zerbi}, \&
  {Zhang}}]{Gehrels2004}
{Gehrels}, N., {Chincarini}, G., {Giommi}, P., {et~al.} 2004, \apj, 611, 1005

\bibitem[{{G\"otz} \& et~al.(2018)}]{Goetz2018a}
{G\"otz}, D. \& et~al. 2018, this prooceeding

\bibitem[{{Greiner} {et~al.}(2011){Greiner}, {Kr{\"u}hler}, {Klose}, {Afonso},
  {Clemens}, {Filgas}, {Hartmann}, {K{\"u}pc{\"u} Yolda{\c s}}, {Nardini},
  {Olivares E.}, {Rau}, {Rossi}, {Schady}, \& {Updike}}]{Greiner2011}
{Greiner}, J., {Kr{\"u}hler}, T., {Klose}, S., {et~al.} 2011, \aap, 526, A30

\bibitem[{{Hjorth} {et~al.}(2012){Hjorth}, {Malesani}, {Jakobsson}, {Jaunsen},
  {Fynbo}, {Gorosabel}, {Kr{\"u}hler}, {Levan}, {Micha{\l}owski},
  {Milvang-Jensen}, {M{\o}ller}, {Schulze}, {Tanvir}, \&
  {Watson}}]{Hjorth2012a}
{Hjorth}, J., {Malesani}, D., {Jakobsson}, P., {et~al.} 2012, \apj, 756, 187

\bibitem[{{Hjorth} {et~al.}(2003){Hjorth}, {Sollerman}, {M{\o}ller}, {Fynbo},
  {Woosley}, {Kouveliotou}, {Tanvir}, {Greiner}, {Andersen}, {Castro-Tirado},
  {Castro Cer{\'o}n}, {Fruchter}, {Gorosabel}, {Jakobsson}, {Kaper}, {Klose},
  {Masetti}, {Pedersen}, {Pedersen}, {Pian}, {Palazzi}, {Rhoads}, {Rol}, {van
  den Heuvel}, {Vreeswijk}, {Watson}, \& {Wijers}}]{Hjorth2003a}
{Hjorth}, J., {Sollerman}, J., {M{\o}ller}, P., {et~al.} 2003, \nat, 423, 847

\bibitem[{{Kann} {et~al.}(2011){Kann}, {Klose}, {Zhang}, {Covino}, {Butler},
  {Malesani}, {Nakar}, {Wilson}, {Antonelli}, {Chincarini}, {Cobb}, {D'Avanzo},
  {D'Elia}, {Della Valle}, {Ferrero}, {Fugazza}, {Gorosabel}, {Israel},
  {Mannucci}, {Piranomonte}, {Schulze}, {Stella}, {Tagliaferri}, \&
  {Wiersema}}]{Kann2011a}
{Kann}, D.~A., {Klose}, S., {Zhang}, B., {et~al.} 2011, \apj, 734, 96

\bibitem[{{Kann} {et~al.}(2010){Kann}, {Klose}, {Zhang}, {Malesani}, {Nakar},
  {Pozanenko}, {Wilson}, {Butler}, {Jakobsson}, {Schulze}, {Andreev},
  {Antonelli}, {Bikmaev}, {Biryukov}, {B{\"o}ttcher}, {Burenin}, {Castro
  Cer{\'o}n}, {Castro-Tirado}, {Chincarini}, {Cobb}, {Covino}, {D'Avanzo},
  {D'Elia}, {Della Valle}, {de Ugarte Postigo}, {Efimov}, {Ferrero}, {Fugazza},
  {Fynbo}, {G{\aa}lfalk}, {Grundahl}, {Gorosabel}, {Gupta}, {Guziy}, {Hafizov},
  {Hjorth}, {Holhjem}, {Ibrahimov}, {Im}, {Israel}, {Je{\'l}inek}, {Jensen},
  {Karimov}, {Khamitov}, {Kizilo{\v g}lu}, {Klunko}, {Kub{\'a}nek}, {Kutyrev},
  {Laursen}, {Levan}, {Mannucci}, {Martin}, {Mescheryakov}, {Mirabal},
  {Norris}, {Ovaldsen}, {Paraficz}, {Pavlenko}, {Piranomonte}, {Rossi},
  {Rumyantsev}, {Salinas}, {Sergeev}, {Sharapov}, {Sollerman}, {Stecklum},
  {Stella}, {Tagliaferri}, {Tanvir}, {Telting}, {Testa}, {Updike}, {Volnova},
  {Watson}, {Wiersema}, \& {Xu}}]{Kann2010a}
{Kann}, D.~A., {Klose}, S., {Zhang}, B., {et~al.} 2010, \apj, 720, 1513

\bibitem[{{Kann} {et~al.}(2017){Kann}, {Schady}, {Olivares E.}, {Klose},
  {Rossi}, {Perley}, {Kr{\"u}hler}, {Greiner}, {Nicuesa Guelbenzu}, {Elliott},
  {Knust}, {Filgas}, {Pian}, {Mazzali}, {Fynbo}, {Leloudas}, {Afonso},
  {Delvaux}, {Graham}, {Rau}, {Schmidl}, {Schulze}, {Tanga}, {Updike}, \&
  {Varela}}]{Kann2017a}
{Kann}, D.~A., {Schady}, P., {Olivares E.}, F., {et~al.} 2017, ArXiv e-prints

\bibitem[{{Levan} {et~al.}(2014){Levan}, {Tanvir}, {Starling}, {Wiersema},
  {Page}, {Perley}, {Schulze}, {Wynn}, {Chornock}, {Hjorth}, {Cenko},
  {Fruchter}, {O'Brien}, {Brown}, {Tunnicliffe}, {Malesani}, {Jakobsson},
  {Watson}, {Berger}, {Bersier}, {Cobb}, {Covino}, {Cucchiara}, {de Ugarte
  Postigo}, {Fox}, {Gal-Yam}, {Goldoni}, {Gorosabel}, {Kaper}, {Kr{\"u}hler},
  {Karjalainen}, {Osborne}, {Pian}, {S{\'a}nchez-Ram{\'{\i}}rez}, {Schmidt},
  {Skillen}, {Tagliaferri}, {Th{\"o}ne}, {Vaduvescu}, {Wijers}, \&
  {Zauderer}}]{Levan2014a}
{Levan}, A.~J., {Tanvir}, N.~R., {Starling}, R.~L.~C., {et~al.} 2014, \apj,
  781, 13

\bibitem[{{Perley} {et~al.}(2017){Perley}, {Kr{\"u}hler}, {Schady},
  {Micha{\l}owski}, {Th{\"o}ne}, {Petry}, {Graham}, {Greiner}, {Klose},
  {Schulze}, \& {Kim}}]{Perley2017a}
{Perley}, D.~A., {Kr{\"u}hler}, T., {Schady}, P., {et~al.} 2017, \mnras, 465,
  L89

\bibitem[{{Perley} {et~al.}(2016){Perley}, {Kr{\"u}hler}, {Schulze}, {de Ugarte
  Postigo}, {Hjorth}, {Berger}, {Cenko}, {Chary}, {Cucchiara}, {Ellis}, {Fong},
  {Fynbo}, {Gorosabel}, {Greiner}, {Jakobsson}, {Kim}, {Laskar}, {Levan},
  {Micha{\l}owski}, {Milvang-Jensen}, {Tanvir}, {Th{\"o}ne}, \&
  {Wiersema}}]{Perley2016a}
{Perley}, D.~A., {Kr{\"u}hler}, T., {Schulze}, S., {et~al.} 2016, \apj, 817, 7

\bibitem[{{Racusin} {et~al.}(2009){Racusin}, {Liang}, {Burrows}, {Falcone},
  {Sakamoto}, {Zhang}, {Zhang}, {Evans}, \& {Osborne}}]{Racusin2009a}
{Racusin}, J.~L., {Liang}, E.~W., {Burrows}, D.~N., {et~al.} 2009, \apj, 698,
  43

\bibitem[{{Roming} {et~al.}(2005){Roming}, {Kennedy}, {Mason}, {Nousek}, {Ahr},
  {Bingham}, {Broos}, {Carter}, {Hancock}, {Huckle}, {Hunsberger}, {Kawakami},
  {Killough}, {Koch}, {McLelland}, {Smith}, {Smith}, {Soto}, {Boyd},
  {Breeveld}, {Holland}, {Ivanushkina}, {Pryzby}, {Still}, \&
  {Stock}}]{Roming2005a}
{Roming}, P.~W.~A., {Kennedy}, T.~E., {Mason}, K.~O., {et~al.} 2005, \ssr, 120,
  95

\bibitem[{{Roming} {et~al.}(2009){Roming}, {Koch}, {Oates}, {Porterfield},
  {Vanden Berk}, {Boyd}, {Holland}, {Hoversten}, {Immler}, {Marshall}, {Page},
  {Racusin}, {Schneider}, {Breeveld}, {Brown}, {Chester}, {Cucchiara},
  {DePasquale}, {Gronwall}, {Hunsberger}, {Kuin}, {Landsman}, {Schady}, \&
  {Still}}]{Roming2009a}
{Roming}, P.~W.~A., {Koch}, T.~S., {Oates}, S.~R., {et~al.} 2009, \apj, 690,
  163

\bibitem[{{Rossi} {et~al.}(2012){Rossi}, {Klose}, {Ferrero}, {Greiner},
  {Arnold}, {Gonsalves}, {Hartmann}, {Updike}, {Kann}, {Kr{\"u}hler},
  {Palazzi}, {Savaglio}, {Schulze}, {Afonso}, {Amati}, {Castro-Tirado},
  {Clemens}, {Filgas}, {Gorosabel}, {Hunt}, {K{\"u}pc{\"u} Yolda{\c s}},
  {Masetti}, {Nardini}, {Nicuesa Guelbenzu}, {Olivares}, {Pian}, {Rau},
  {Schady}, {Schmidl}, {Yolda{\c s}}, \& {de Ugarte Postigo}}]{Rossi2012a}
{Rossi}, A., {Klose}, S., {Ferrero}, P., {et~al.} 2012, \aap, 545, A77

\bibitem[{{Rossi} {et~al.}(2014){Rossi}, {Piranomonte}, {Savaglio}, {Palazzi},
  {Micha{\l}owski}, {Klose}, {Hunt}, {Amati}, {Elliott}, {Greiner}, {Guidorzi},
  {Japelj}, {Kann}, {Lo Faro}, {Nicuesa Guelbenzu}, {Schulze}, {Vergani},
  {Arnold}, {Covino}, {D'Elia}, {Ferrero}, {Filgas}, {Goldoni}, {K{\"u}pc{\"u}
  Yolda{\c s}}, {Le Borgne}, {Pian}, {Schady}, \& {Stratta}}]{Rossi2014a}
{Rossi}, A., {Piranomonte}, S., {Savaglio}, S., {et~al.} 2014, \aap, 572, A47

\bibitem[{{Rossi} {et~al.}(2011){Rossi}, {Schulze}, {Klose}, {Kann}, {Rau},
  {Krimm}, {J{\'o}hannesson}, {Panaitescu}, {Yuan}, {Ferrero}, {Kr{\"u}hler},
  {Greiner}, {Schady}, {Pandey}, {Amati}, {Afonso}, {Akerlof}, {Arnold},
  {Clemens}, {Filgas}, {Hartmann}, {K{\"u}pc{\"u} Yolda{\c s}}, {McBreen},
  {McKay}, {Nicuesa Guelbenzu}, {Olivares}, {Paciesas}, {Rykoff}, {Szokoly},
  {Updike}, \& {Yolda{\c s}}}]{Rossi2011a}
{Rossi}, A., {Schulze}, S., {Klose}, S., {et~al.} 2011, \aap, 529, A142

\bibitem[{{Salvaterra} {et~al.}(2012){Salvaterra}, {Campana}, {Vergani},
  {Covino}, {D'Avanzo}, {Fugazza}, {Ghirlanda}, {Ghisellini}, {Melandri},
  {Nava}, {Sbarufatti}, {Flores}, {Piranomonte}, \&
  {Tagliaferri}}]{Salvaterra2012a}
{Salvaterra}, R., {Campana}, S., {Vergani}, S.~D., {et~al.} 2012, \apj, 749, 68

\bibitem[{{Schady} {et~al.}(2012){Schady}, {Dwelly}, {Page}, {Kr{\"u}hler},
  {Greiner}, {Oates}, {de Pasquale}, {Nardini}, {Roming}, {Rossi}, \&
  {Still}}]{Schady2012a}
{Schady}, P., {Dwelly}, T., {Page}, M.~J., {et~al.} 2012, \aap, 537, A15

\bibitem[{{Schulze} {et~al.}(2011){Schulze}, {Klose}, {Bj{\"o}rnsson},
  {Jakobsson}, {Kann}, {Rossi}, {Kr{\"u}hler}, {Greiner}, \&
  {Ferrero}}]{Schulze2011}
{Schulze}, S., {Klose}, S., {Bj{\"o}rnsson}, G., {et~al.} 2011, \aap, 526, A23

\bibitem[{{Stratta} {et~al.}(2013){Stratta}, {Gendre}, {Atteia}, {Bo{\"e}r},
  {Coward}, {De Pasquale}, {Howell}, {Klotz}, {Oates}, \&
  {Piro}}]{Stratta2013a}
{Stratta}, G., {Gendre}, B., {Atteia}, J.~L., {et~al.} 2013, \apj, 779, 66

\bibitem[{{Troja} {et~al.}(2017){Troja}, {Lipunov}, {Mundell}, {Butler},
  {Watson}, {Kobayashi}, {Cenko}, {Marshall}, {Ricci}, {Fruchter}, {Wieringa},
  {Gorbovskoy}, {Kornilov}, {Kutyrev}, {Lee}, {Toy}, {Tyurina}, {Budnev},
  {Buckley}, {Gonz{\'a}lez}, {Gress}, {Horesh}, {Panasyuk}, {Prochaska},
  {Ramirez-Ruiz}, {Rebolo Lopez}, {Richer}, {Roman-Zuniga}, {Serra-Ricart},
  {Yurkov}, \& {Gehrels}}]{Troja2017a}
{Troja}, E., {Lipunov}, V.~M., {Mundell}, C.~G., {et~al.} 2017, \nat, 547, 425

\bibitem[{{Zhang} \& {M{\'e}sz{\'a}ros}(2001)}]{ZhangMeszaros2001a}
{Zhang}, B. \& {M{\'e}sz{\'a}ros}, P. 2001, \apjl, 552, L35

\end{thebibliography}

\end{document}